\documentclass[aps,pra,amsmath,twocolumn,showpacs,superscriptaddress]{revtex4-1}
\usepackage[english]{babel}
\usepackage{graphicx}
\usepackage{subfigure}
\usepackage{dcolumn}
\usepackage{bm}
\usepackage{units}
\usepackage[colorlinks=true,linkcolor=blue,citecolor=blue,urlcolor=blue,filecolor=blue]{hyperref}
\usepackage{color}
\usepackage{xcolor}
\usepackage{longtable}
\usepackage{array}
\usepackage{multirow}
\usepackage{physics}

\newcommand{\beginmethods}{
        \setcounter{table}{0}
        \renewcommand{\thetable}{M\arabic{table}}
        \setcounter{figure}{0}
        \renewcommand{\thefigure}{M\arabic{figure}}
     }

\newcolumntype{L}[1]{>{\raggedright\let\newline\\\arraybackslash\hspace{0pt}}m{#1}}
\newcolumntype{C}[1]{>{\centering\let\newline\\\arraybackslash\hspace{0pt}}m{#1}}
\newcolumntype{R}[1]{>{\raggedleft\let\newline\\\arraybackslash\hspace{0pt}}m{#1}}
\newcommand{\rem}[1]{\ignorespaces}

\begin{document}

\title{Observation of Feshbach resonances between alkali and closed-shell atoms}
\author{Vincent Barb\'{e}}
\email[]{RbSrFR@strontiumBEC.com}
\thanks{\newline These authors contributed equally to this work.}
\affiliation{Van der Waals-Zeeman Institute, Institute of Physics, University of Amsterdam, Science Park 904, 1098XH Amsterdam, The Netherlands}
\author{Alessio Ciamei}
\email[]{RbSrFR@strontiumBEC.com}
\thanks{\newline These authors contributed equally to this work.}
\affiliation{Van der Waals-Zeeman Institute, Institute of Physics, University of Amsterdam, Science Park 904, 1098XH Amsterdam, The Netherlands}
\author{Benjamin Pasquiou}
\affiliation{Van der Waals-Zeeman Institute, Institute of Physics, University of Amsterdam, Science Park 904, 1098XH Amsterdam, The Netherlands}
\author{Lukas Reichs\"{o}llner}
\affiliation{Van der Waals-Zeeman Institute, Institute of Physics, University of Amsterdam, Science Park 904, 1098XH Amsterdam, The Netherlands}
\author{Florian Schreck}
\affiliation{Van der Waals-Zeeman Institute, Institute of Physics, University of Amsterdam, Science Park 904, 1098XH Amsterdam, The Netherlands}
\author{Piotr S. \ifmmode \dot{Z}\else \.{Z}\fi{}uchowski}
\affiliation{Institute of Physics, Faculty of Physics, Astronomy and
Informatics, Nicolaus Copernicus University, ul.\ Grudziadzka 5/7, 87-100
Torun, Poland}
\author{Jeremy M. Hutson}
\affiliation{Joint Quantum Centre (JQC) Durham-Newcastle, Department of
Chemistry, Durham University, South Road, Durham, DH1 3LE, United Kingdom}

\date{\today}

\maketitle

\textbf{Magnetic Feshbach resonances are an invaluable tool for controlling
ultracold atoms and molecules \cite{Chin2010fri}. They can be used to tune
atomic interactions and have been used extensively to explore few- and
many-body phenomena \cite{Greene2017ufb, Bloch2008mbp}. They can also be used
for magnetoassociation, in which pairs of atoms are converted into molecules by
ramping an applied magnetic field across a resonance \cite{Hutson:IRPC:2006,
Kohler2006poc}. Pairs of open-shell atoms, such as the alkalis, chromium
\cite{Werner:2005}, and some lanthanides \cite{Frisch2014qci, Baumann:2014,
Taie:FermYb2:2016}, exhibit broad resonances because the corresponding molecule
has multiple electronic states. However, molecules formed between alkali and
closed-shell atoms have only one electronic state and no broad resonances.
Narrow resonances have been predicted in such systems \cite{Zuchowski2010urm,
Brue2012mtf, Brue2013pof}, but until now have eluded observation.  Here we
present the first observation of magnetic Feshbach resonances in a system
containing a closed-shell atom, Sr, interacting with an alkali atom, Rb. These
resonances pave the way to creating an ultracold gas of strongly polar,
open-shell molecules, which will open up new possibilities for designing
quantum many-body systems \cite{Micheli2006atf, Baranov2012cmt} and for tests
of fundamental symmetries \cite{Meyer2009eed}.}

A magnetic Feshbach resonance arises when a pair of ultracold atoms couples to
a near-threshold molecular state that is tuned to be close in energy by an
applied magnetic field. Magnetoassociation at such a resonance
coherently transfers the atoms into the molecular state \cite{Regal:40K2:2003,
Herbig:2003}. In a few cases, near-threshold molecules formed in this way
have been transferred to their absolute ground states \cite{Ni:KRb:2008,
Danzl:ground:2010, Takekoshi:RbCs:2014}, allowing exploration of quantum gases
with strong dipolar interactions \cite{Moses:2017}.  However, this
has so far been achieved only for molecules formed from pairs of alkali atoms.

Mixtures of closed-shell alkaline-earth atoms with open-shell alkali atoms have
been studied in several laboratories \cite{Hara2011qdm, Hansen2011qdm,
Borkowski2013sli, Guttridge:2017}. No strong coupling mechanism between atomic
and molecular states exists in systems of this type, but theoretical work has
identified weak coupling mechanisms that should lead to narrow Feshbach
resonances, suitable for magnetoassociation \cite{Zuchowski2010urm,
Brue2012mtf, Brue2013pof}. In this letter we describe the detection of Feshbach
resonances in mixtures of $^{87}$Sr or $^{88}$Sr with $^{87}$Rb. The coupling
between atomic and molecular states arises from two mechanisms previously
predicted \cite{Zuchowski2010urm, Brue2012mtf, Brue2013pof} and an additional,
weaker mechanism that we identify here. The energies of the bound states
responsible for the resonances are confirmed by two-photon photoassociation
spectroscopy.

\begin{figure*}[tb]
\centering
\subfigure{\includegraphics[width=.8\textwidth]{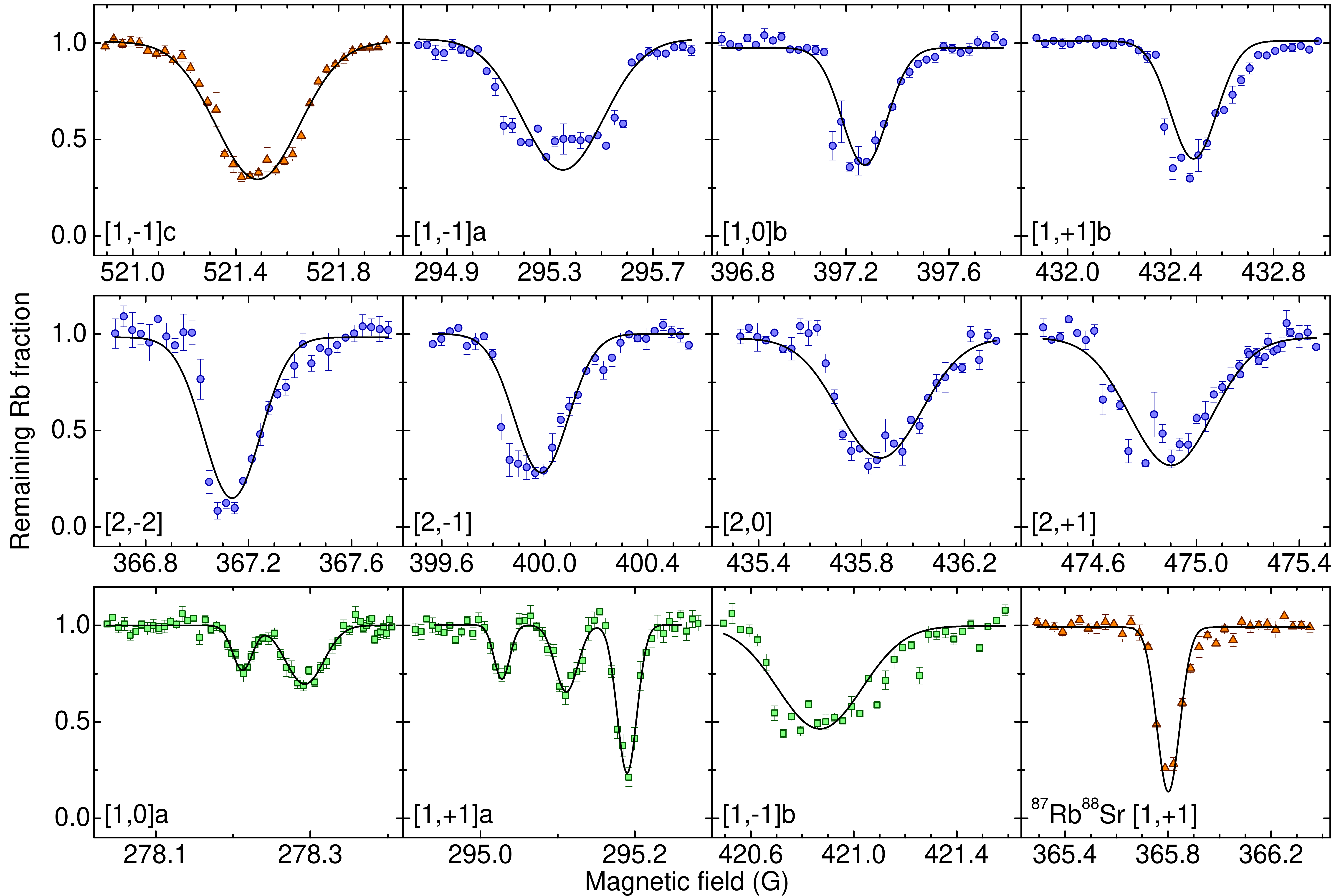}}
\caption{\label{fig:Fig1_LossFeatures} \textbf{Detection of Rb-Sr Feshbach
resonances by field-dependent loss of Rb.} The fraction of Rb atoms remaining
in state $(f,m_f)$ after loss at each observed Feshbach resonance, normalised
to unity far from the loss feature. Eleven loss features are observed in
$^{87}$Rb\,-$^{87}$Sr mixtures and one in $^{87}$Rb\,-$^{88}$Sr (lower right
panel). The loss features are labelled by [$f$,$m_f$]$j$, where $j\in
$\{a,b,c\} is an index used when losses due to several molecular states are
observed at the same atomic threshold. Most loss features show a single dip in
the atom number, whereas [1,0]a and [1,1]a show several. Each dip is fit by a
Gaussian (black line), with results shown in Tab.\,\ref{tab:FRList}. The color
and shape of symbols indicates the coupling mechanism for the Feshbach
resonance: mechanism I (orange triangles), II (blue circles), or III (green
squares). The resonance near 521\,G also has a contribution from mechanism II.
Error bars represent the standard error of three or more data points.}
\end{figure*}

The experimental signature of a Feshbach resonance is field-dependent loss of
Rb atoms. This may arise from either 3-body recombination or inelastic
collisions, both of which are enhanced near a resonance. We perform loss
spectroscopy using an ultracold Rb-Sr mixture, typically consisting of $5\times
10^4$ Rb atoms mixed with $10^6$ $^{87}$Sr or $10^7$ $^{88}$Sr atoms at a
temperature of 2 to $\unit[5]\mu$K (see Methods). Figure
\ref{fig:Fig1_LossFeatures} shows the observed loss features, eleven arising in
the $^{87}$Rb\,-$^{87}$Sr Bose-Fermi mixture and one in the
$^{87}$Rb\,-$^{88}$Sr Bose-Bose mixture. Ten loss features consist of a single,
slightly asymmetrical dip with FWHM between 200 and $\unit[400]{mG}$. The loss
features labelled [1,0]a and [1,1]a each consist of several dips with a width
of 20 to $\unit[60]{mG}$ at a spacing of $\unit[80]{mG}$. We fit each dip with
a Gaussian and give the resulting positions and widths in
Tab.\,\ref{tab:FRList}. None of these Rb loss features arises in the absence of
Sr, proving that they depend on Rb-Sr interactions. We also observe Rb loss
features in the absence of Sr, which coincide with known Rb Feshbach resonances
\cite{Marte2002fri}.

Both the atomic and molecular states are described by the total angular
momentum of the Rb atom, $f$, and its projection $m_f$ onto the magnetic field.
Where necessary, atomic and molecular quantum numbers are distinguished with
subscripts at and mol. In addition, the molecule has a vibrational quantum
number $n$, counted down from $n=-1$ for the uppermost level, and a rotational
quantum number $L$, with projection $M_L$. $^{88}$Sr has nuclear spin $i_{\rm
Sr}=0$, whereas $^{87}$Sr has $i_{\rm Sr}=9/2$ and a corresponding projection
$m_{i,{\rm Sr}}$.

\begin{figure}[t]
\centering
\subfigure{\includegraphics[width=.98\columnwidth]{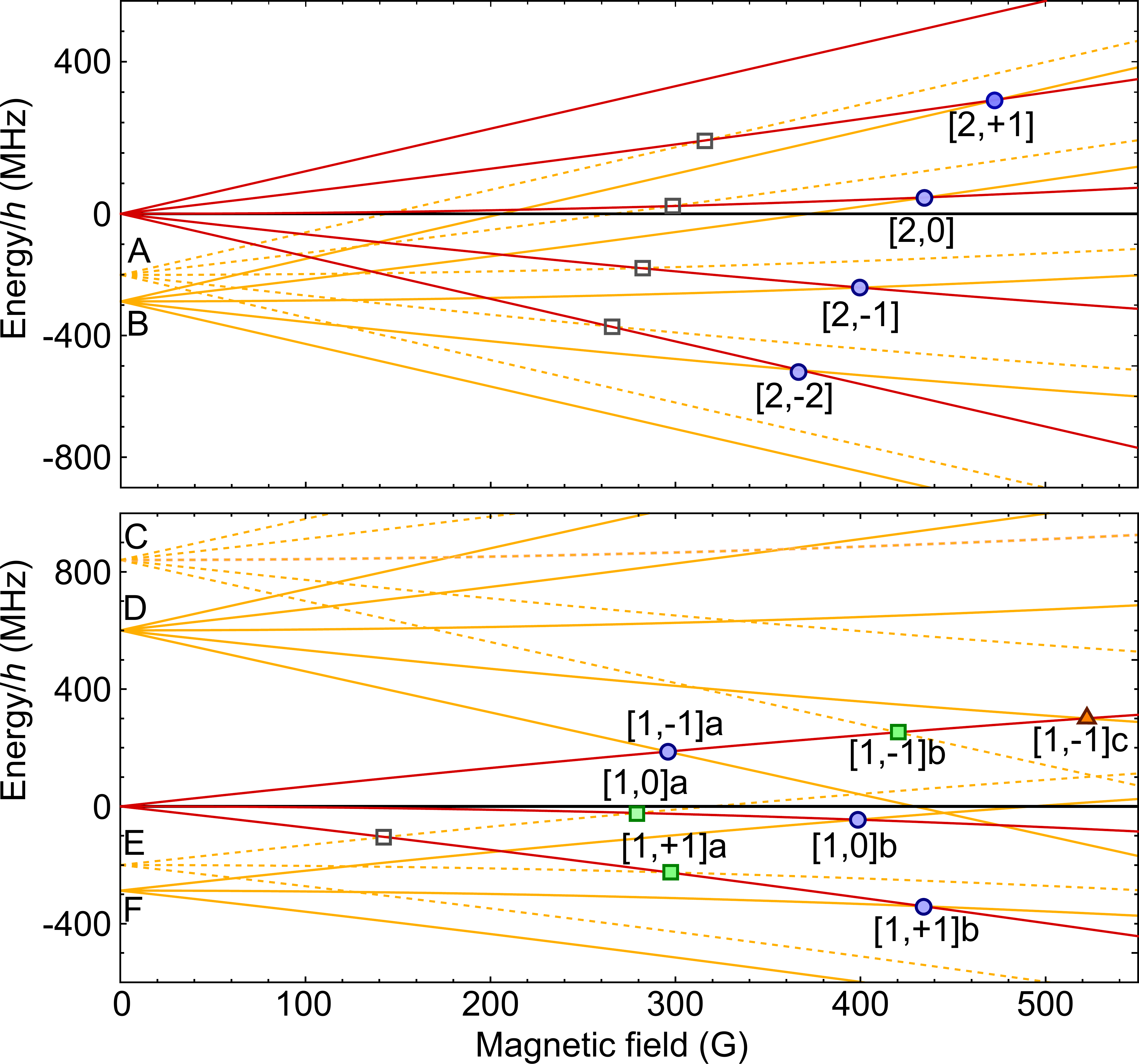}}

\caption{\label{fig:Fig2_OverviewRb87Sr87} \textbf{Origin of the
$^{87}$Rb\,-$^{87}$Sr Feshbach resonances.} Energies of atomic (red) and
molecular (orange) states as functions of magnetic field, shown with respect to
the zero-field atomic level with $f=1$ or 2 as appropriate. Molecular states
are labelled as in Tab.\,\ref{tab:BoundStates} and shown dashed if rotationally
excited ($L=2)$. Observed Feshbach resonances are labelled as in
Fig.\,\ref{fig:Fig1_LossFeatures} and marked by filled symbols (orange
triangles, blue circles or green squares for coupling mechanism I, II or III,
respectively). Predicted but unobserved Feshbach resonances are marked by
hollow symbols.}
\end{figure}

\begin{figure}[t]
\centering
\subfigure{\includegraphics[width=.98\columnwidth]{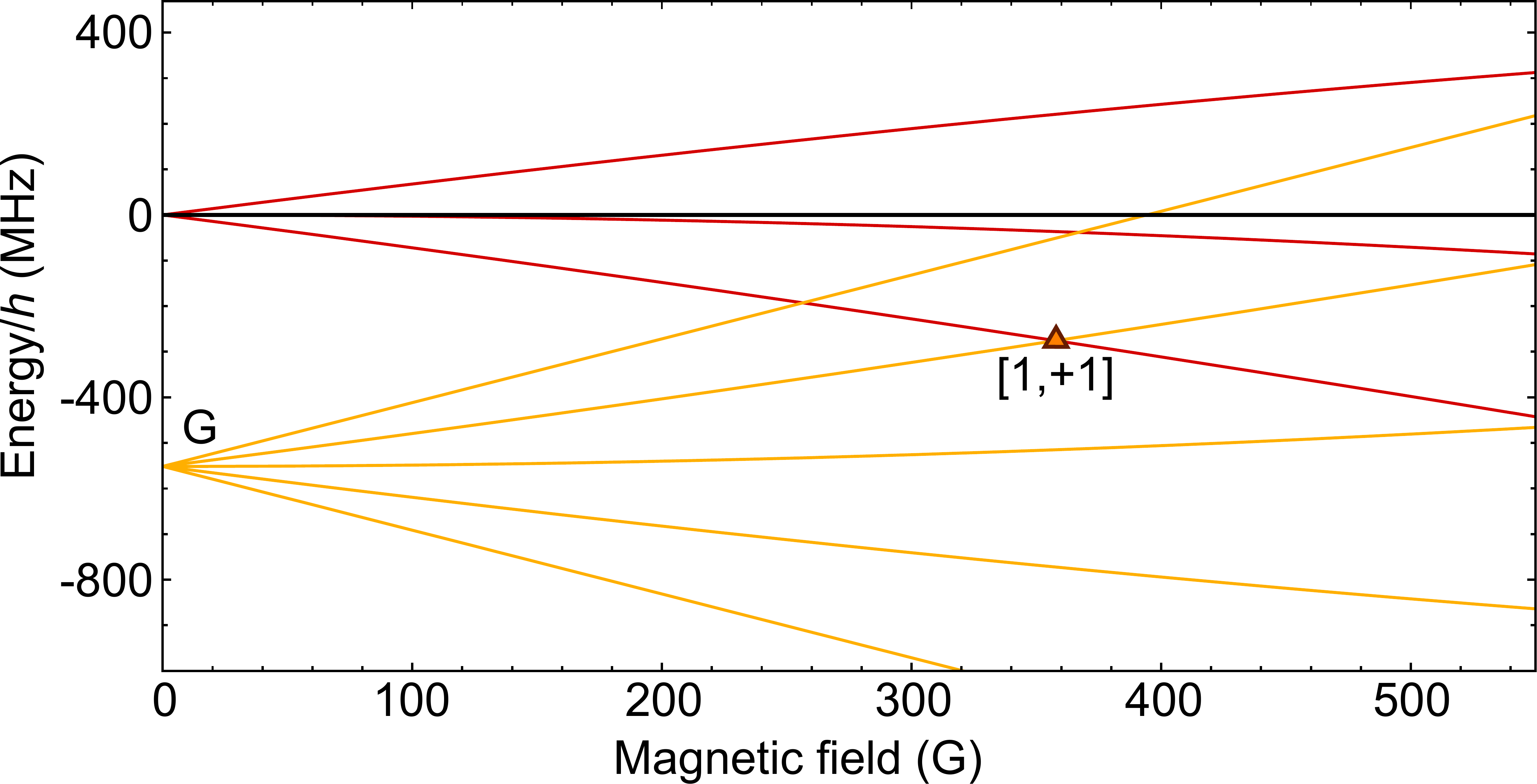}}

\caption{\label{fig:Fig3_OverviewRb87Sr88} \textbf{Origin of the
$^{87}$Rb\,-$^{88}$Sr Feshbach resonance.} Energies of atomic (red) and
molecular (orange) states as functions of magnetic field, shown with respect to
the zero-field $f=1$ atomic level. Only one Feshbach resonance has been
observed, produced by coupling mechanism I. Since $^{88}$Sr has zero nuclear
spin, mechanism II is absent.}
\end{figure}

The near-threshold molecular states lie almost parallel to the Rb atomic states
as a function of magnetic field. This is because the presence of the Sr atom
barely changes the Rb hyperfine structure, and the Sr hyperfine energy is very
small. We can therefore use the Breit-Rabi formula for Rb to convert the
resonance positions into zero-field binding energies $E_{\rm b}$ for the
molecular states, which are given in Tab.\,\ref{tab:BoundStates}. The crossing
atomic and molecular levels are shown in Figs.\,\ref{fig:Fig2_OverviewRb87Sr87}
and \ref{fig:Fig3_OverviewRb87Sr88}, with filled symbols where we observe loss
features.

To verify the bound-state energies and validate our model of Feshbach
resonances, we use two-photon photoassociation (PA) spectroscopy. We detect the
two $n=-2$ states (with $L=0$ and 2) below the lower ($f=1$) threshold of
$^{87}$Rb-$^{87}$Sr (states E and F in Tab.\,\ref{tab:BoundStates}) at almost
exactly the energies deduced from the resonance positions. All the states
observed through Feshbach resonances (B to F) also arise to within
$\unit[2]{MHz}$ in a more complete model of the Rb-Sr interaction potential, as
described below.

Three different coupling mechanisms are responsible for the observed loss
features. The first mechanism was proposed in ref.\ \cite{Zuchowski2010urm} and
relies on the change of the Rb hyperfine splitting when the Rb electron
distribution is perturbed by an approaching Sr atom. Its coupling
strength is proportional to the magnetic field in the field region explored
here \cite{Brue2013pof}. Since only states of equal $m_f$ and $L$ are coupled,
it leads to Feshbach resonances only at crossings between atomic states with Rb
in $f=1$ and molecular states with $L=0$ that correlate with $f=2$. We observe
one such resonance with each of $^{87}$Sr and $^{88}$Sr.

The second mechanism involves hyperfine coupling of the Sr nucleus to the
valence electron of Rb and was first proposed in ref.\ \cite{Brue2012mtf}.
Since only fermionic $^{87}$Sr has a nuclear magnetic moment, this can occur
only in Rb\,-$^{87}$Sr collisions. This coupling conserves $L$ and
$m_f+m_{i,{\rm Sr}}$, with the selection rule $m_{f,{\rm at}}-m_{f,{\rm mol}} =
0, \pm 1$. Crossings that fulfil these conditions occur also for molecular
states with the same $f$ value as the atomic state, which makes them much more
abundant than crossings obeying the selection rules of the first mechanism.
Feshbach resonances belonging to different $m_{i,{\rm Sr}}$ are slightly
shifted with respect to one another because of the weak Zeeman effect on the Sr
nucleus and the weak Sr hyperfine splitting. However, since the shift is only
$\sim \unit[10]{mG}$ for neighboring $m_{i,{\rm Sr}}$, much smaller than the
width of the loss features of typically $\unit[300]{mG}$, we do not resolve
this splitting.

The third mechanism is the anisotropic interaction of the electron spin with
the nucleus of either Rb or fermionic Sr. This mechanism can couple the
$s$-wave atomic state to molecules with rotational quantum number $L=2$. As
usual, the total angular momentum projection (now $m_f+m_{i,{\rm Sr}}+M_L$) is
conserved. If the Sr nucleus is involved, an additional selection rule is
$\Delta m_f=\pm 1$. By contrast, if the Rb nucleus is involved, the selection
rule is $\Delta m_f=-\Delta M_L$. These loss features are made up of many
$(m_f,M_L)$ components, split by several hyperfine terms
\cite{Aldegunde:doublet:2017}; in some cases the components separate into
groups for different values of $M_L$. Three loss features are attributed to
this mechanism and two of them ([1,1]a and [1,0]a) indeed show a structure of
two or three dips.

Table\ \ref{tab:FRList} includes a theoretical width $\Delta$, obtained from
the Golden Rule approximation \cite{Brue2013pof}. However, this is a physically
different quantity from the experimental width $\delta$, and for narrow
resonances there is no simple link between them. We have also
searched for further resonances predicted by our model, marked by hollow
symbols in Fig.\,\ref{fig:Fig2_OverviewRb87Sr87}, but did not observe them.

\begin{table}[b]

\caption{\textbf{Properties of observed Feshbach resonances.} For resonances
with many components, the theoretical width is the largest calculated value.}
\begin{ruledtabular}
\begin{tabular}{llcccl}
  \noalign{\smallskip}
$[f, m_f]j$ & (mol. state,  & $B$ & $\delta$   & $\Delta$  & cpl. \\
  &  $m_f$, $M_L$) & (G) & (mG)  & (mG) &  mech. \\
  \noalign{\smallskip}
  \hline
  \noalign{\medskip}
  \noalign{$^{87}$Rb\,-$^{87}$Sr}
  \noalign{\medskip}
  \hline
  \noalign{\smallskip}
{[2,$+1$]} & (B, $+2$, 0) & 474.9(4) & 373(7) & 0.04 & II \\
{[2,$0$]} & (B, $+1$, 0) & 435.9(4) & 378(7) & 0.07 & II \\
{[2,$-1$]} & (B, 0, 0) & 400.0(4) & 247(4) & 0.07 & II \\
{[2,$-2$]} & (B, $-1$, 0) & 367.1(4) & 260(5) & 0.04 & II \\
{[1,$-1$]a} & (D, $-2$, 0) & 295.4(4) & 372(10) & 0.33 & II \\
{[1,$-1$]b} & (C, $-2$, {\rm mix}) & 420.9(4) & 386(11) & 0.002 & III \\
{[1,$-1$]c} & (D, $-1$, 0) & 521.5(4)  & 366(3) & 2.4 & I,II \\
{[1,$0$]a} & (E, $-1$, $-1$) & $B_{\rm 1}=278.2(4)$  & 30(3) & 0.00009 & III \\
            & (E, $-1$, $-2$) & $B_{\rm 1}+0.081(2)$  & 58(4) & 0.00011 & III \\
{[1,$0$]b} & (F, $-1$, 0) & 397.3(4)  & 207(4) & 0.02 & II \\
{[1,$+1$]a} & (E, $0$, 0) & $B_{\rm 2}= 295.0(4)$  & 24(3) & 0.00002 & III \\
            & (E, $0$, $-1$) & $B_{\rm 2}+ 0.083(2)$  & 35(3) & 0.00009 & III \\
            & (E, $0$, $-2$) & $B_{\rm 2}+ 0.162(2)$  & 30(1) & 0.00011 & III \\
{[1,$+1$]b} & (F, 0, 0) & 432.5(4)  & 213(6)  & 0.02 & II \\
  \noalign{\smallskip}
  \hline
  \noalign{\medskip}
  \noalign{$^{87}$Rb\,-$^{88}$Sr}
  \noalign{\medskip}
  \hline
  \noalign{\smallskip}
{[1,$+1$]} & (G, $+1$, 0) & 365.8(4)  & 105(2) & 0.05 & I \\
 \noalign{\smallskip}
\end{tabular}
\end{ruledtabular}
\label{tab:FRList} \end{table}

\begin{table}[b]
\caption{\textbf{Molecular states responsible for Feshbach
resonances.} Binding energies obtained from observed Feshbach resonances,
$E_\textrm{b}^\textrm{FR}$, and from two-photon photoassociation,
$E_\textrm{b}^\textrm{PA}$.}
\begin{ruledtabular}
\begin{tabular}{ccccrr}
  \noalign{\smallskip}
label & $n$ & $F$ & $L$  & $E_\textrm{b}^\textrm{FR}/h$   & $E_\textrm{b}^{\rm PA}/h$    \\
  &   &   &    &  (MHz) &   (MHz)  \\
  \noalign{\smallskip}
  \hline
  \noalign{\medskip}
  \noalign{$^{87}$Rb\,-$^{87}$Sr}
  \noalign{\medskip}
  \hline
  \noalign{\smallskip}
A & $-2$ & 2 & 2 & - & - \\
B & $-2$ & 2 & 0 & 288.2(4) & - \\
C & $-4$ & 2 & 2 & 5992(1) & - \\
D & $-4$ & 2 & 0 & 6234(1) & - \\
E & $-2$ & 1 & 2 & 200.0(3) & 200.0(3) \\
F & $-2$ & 1 & 0 & 287.3(3) & 287.3(2) \\
\noalign{\smallskip}
  \hline
  \noalign{\medskip}
  \noalign{$^{87}$Rb\,-$^{88}$Sr}
  \noalign{\medskip}
  \hline
  \noalign{\smallskip}
G & $-4$ & 2 & 0 & 7401.0(7) & - \\
 \noalign{\smallskip}
\end{tabular}
\end{ruledtabular}
\label{tab:BoundStates}
\end{table}

We have previously carried out electronic structure calculations of the RbSr
ground-state potential \cite{Zuchowski2014gae}. We have used the binding
energies from two-photon photoassociation, supplemented by the Feshbach
resonance positions measured here, to determine a short-range correction to
this potential \cite{GroundStatePotentialArticle}. This allows us to estimate
that the interspecies scattering length of $^{87}$Rb\,-$^{87}$Sr is $a_{\rm
87,87}>\unit[1600(+600,-450)]{a_0}$ and that of $^{87}$Rb\,-$^{88}$Sr is
$a_{\rm 87,88}=\unit[170(20)]{a_0}$, where $a_0$ is the Bohr radius. The large
positive scattering length for $^{87}$Rb\,-$^{87}$Sr will produce a molecular
state with binding energy $h \times \unit[25(15)]{kHz}$, which would lead to
Feshbach resonances at low magnetic field. We searched for such resonances
between $\unit[10]{mG}$ and $\unit[1]{G}$, but did not find any.

There are several factors that affect resonance widths and hence observability.
First, the amplitude of the atomic scattering function at short range depends
on the background scattering length $a$; it is largest when $a$ is large, and
the resonance widths are proportional to $a$ in this regime \cite{Brue2013pof}.
This effect enhances all the resonance widths for $^{87}$Rb\,-$^{87}$Sr.
However, bound states very near dissociation exist mostly at long range, and
the widths also depend on the binding energy as $|E_{\rm b}|^{2/3}$
\cite{Brue2013pof}. This latter effect may explain our failure to observe the
low-field resonances due to the state at $E_{\rm b}\approx h \times
\unit[25(15)]{kHz}$.

Our model enables us to predict the background scattering lengths and Feshbach
resonance positions for all isotopic Rb-Sr mixtures
\cite{GroundStatePotentialArticle}. For example, we predicted the position of
the $^{87}$Rb\,-$^{88}$Sr resonance after calibrating the model on
$^{87}$Rb\,-$^{87}$Sr Feshbach resonances and photoassociation results for
three isotopic mixtures. This resonance was subsequently observed within
$\unit[10]{G}$ of the prediction.

In summary, we have observed Feshbach resonances in mixtures of Rb alkali and
Sr alkaline-earth atoms. Similar resonances will be ubiquitous in mixtures of
alkali atoms with closed-shell atoms, particularly when the closed-shell atom
has a nuclear spin. Magnetoassociation using resonances of this type offers a
path towards a new class of ultracold molecules, with electron spin and strong
electric dipole moment, which are expected to have important applications in
quantum computation, many-body physics and tests of fundamental symmetries.

\begin{acknowledgments}
We thank the European Research Council (ERC) for funding under Project No.\
615117 QuantStro. B.P. thanks the NWO for funding through Veni grant No.\
680-47-438. P.S.\.Z. thanks the Foundation for Polish Science for funding of
Homing Plus project No.\ 2011-3/14 (co-financed by the EU European Regional
Development Fund). J.M.H. thanks the UK Engineering and Physical Sciences Research
Council for support under Grant No.\ EP/P01058X/1.
\end{acknowledgments}

\beginmethods

\section{Methods}

\textbf{Sample preparation.} We prepare ultracold $^{87}$Rb\,-$^{87}$Sr
Bose-Fermi mixtures by methods similar to those in our previous work
\cite{Pasquiou2013qdm}. We transfer Rb and $^{88}$Sr from magneto-optical traps
into a horizontal ``reservoir'' dipole trap with a waist of 63(2)\,$\mu$m and a
wavelength of 1070\,nm. After Rb laser cooling we optically pump Rb into the
$f=1$ hyperfine state. By laser cooling Sr in the dipole trap on the narrow
$^1$S$_0$-$^3$P$_1$ line we sympathetically cool Rb. We then transfer between
$5\times 10^4$ and $1\times 10^5$ Rb atoms into the crossed-beam ``science''
dipole trap described below. We then ramp off the reservoir trap, discard
$^{88}$Sr atoms and transfer between $1 \times 10^6$ and $2 \times 10^6$
$^{87}$Sr atoms in a mixture of all ten nuclear spin states into the science
trap. The final temperature is typically 2 to 5\,$\mu$K. In order to prepare Rb
in an equal mixture of all three $f = 1$ $m_f$ states we then randomize the
distribution by non-adiabatic radiofrequency sweeps at a magnetic field of
130\,G. To prepare Rb in the $f = 2$ hyperfine states we instead use optical
pumping, which directly produces a nearly homogeneous distribution of Rb over
the $f = 2$ $m_f$ states. To prepare $^{87}$Rb\,-$^{88}$Sr Bose-Bose mixtures
we do not discard $^{88}$Sr after transferring the gas into the science trap
and we skip the loading of $^{87}$Sr.

\textbf{Science dipole trap.} The science trap consists of two copropagating
horizontal beams and one vertical beam, all with coinciding foci. The first
horizontal beam has a wavelength of $\unit[1064]{nm}$ and a waist of
$\unit[313(16)]{\mu m}$ ($\unit[19(1)]{\mu m}$) in the horizontal (vertical)
direction. The second horizontal beam has a wavelength of $\unit[532]{nm}$ and
a waist of $\unit[219(4)]{\mu m}$ ($\unit[19(1)]{\mu m}$). The vertical beam
has a wavelength of $\unit[1070]{nm}$ and a waist of $\unit[78(2)]{\mu m}$. The
horizontal 1064-nm beam is typically used at a power of $\unit[5.7]{W}$ to
6.2\,W and dominates the trap potential. The 532-nm beam is operated at 0.2\,W
to 0.4\,W. The vertical beam is operated at $\unit[0.7(1)]{W}$ to measure loss
feature [1,$-1$]b and is off otherwise. These operating conditions result in
typical trap depths of 40\,$\mu$K$\times k_{\rm B}$ for Sr and 95\,$\mu$K$\times k_{\rm B}$ for Rb, taking account of gravitational sag.

\textbf{Loss spectroscopy.} We observe Feshbach resonances through
field-dependent loss of Rb atoms. We submit the Rb-Sr mixture to a magnetic
field of up to $\unit[550]{G}$ for a hold time of 1 to $\unit[10]{s}$. Close to
a Feshbach resonance, the rate of 2-body inelastic collisions or 3-body
recombination is increased and atoms are lost. After the hold time we lower the
magnetic field to near zero in $\unit[200]{ms}$. During the next 10\,ms, we
ramp off the horizontal 532-nm beam and the vertical beam and decrease the
power of the horizontal 1064-nm beam. This decrease lowers the evaporation
threshold for Sr significantly, while Rb stays well trapped because its
polarizability at 1064 nm is a factor of three higher. During the next 100\,ms,
Sr evaporates and cools Rb, which is advantageous for the subsequent imaging
process.

At the end of the cooling stage, the science trap is switched off and a
magnetic field gradient is applied to perform Stern-Gerlach separation of the
Rb $m_f$ states. After $\unit[14]{ms}$ of expansion, absorption images of Sr
and Rb are taken. To reduce sensitivity to Rb atom number fluctuations, we
normalize the atom number of the Rb $m_f$ state of interest by the total atom
number in $m_f$ states that are not lost. We verify that none of the loss
features occurs in the absence of the Sr isotope concerned. For these
verifications we need to retain a small amount of the other Sr isotope to allow
sympathetic cooling.

\textbf{Adaptation of experimental conditions.} The width and depth of a loss
feature depend on the measurement conditions. Thermal broadening sets a lower
bound on the observable width, whereas hold time and peak densities affect the
depth. For each resonance, we optimize the Sr density and hold time to maximize
Rb loss without saturating the feature. We choose to use mixtures that contain
much less Rb than Sr in order to obtain pronounced Rb loss features. Because of
this atom number imbalance, and since most $m_{i,{\rm Sr}}$ states can
contribute to a given loss feature, the fractional Sr loss during the hold time
is small, which keeps the Rb loss rate high.

The resonances that we attribute to coupling mechanism II, with the exception
of [1,$-1$]a, are recorded under identical conditions. The hold time is 5\,s and
the peak densities for Rb and $^{87}$Sr are $2(1)\times 10^{11}\,$cm$^{-3}$ and $9(5)\times 10^{11}\,$cm$^{-3}$,
respectively. The temperature of the Rb-Sr mixture before the hold time is 4.5(5)\,$\mu$K.

The [1,$-1$]a and [1,$-1$]c resonances exhibit higher loss rates.
For these we use hold times of 1.5 and 1\,s, respectively. The peak densities
for Rb and $^{87}$Sr are $5(3)\times 10^{11}\,$cm$^{-3}$ and $2(1)\times 10^{12}\,$cm$^{-3}$.
The temperature of the Rb-Sr mixture before the hold time is 3.0(1)\,$\mu$K.

The resonances that we attribute to mechanism III exhibit much lower loss
rates. Therefore we use a hold time of 10\,s. To measure feature [1,$-1$]b we
also add the vertical trapping beam to increase the gas densities. The peak
densities for Rb are $2(1)\times 10^{12}\,$cm$^{-3}$, $4(2)\times 10^{11}\,$cm$^{-3}$ and $3(2)\times 10^{11}\,$cm$^{-3}$ for features [1,$-1$]b, [1,0]a and [1,+1]a respectively. The peak
densities for Sr are $5(2)\times 10^{12}\,$cm$^{-3}$, $2(1)\times 10^{12}\,$cm$^{-3}$ and $4(2)\times 10^{12}\,$cm$^{-3}$ respectively.
The temperatures of the Rb-Sr mixture before the hold time are respectively 5(1)\,$\mu$K, 3.2(2)\,$\mu$K and 3.0(2)\,$\mu$K.

The [1,+1] resonance observed in the Rb-$^{88}$Sr mixture exhibits a high loss
rate, because we typically load one order of magnitude more $^{88}$Sr atoms
than $^{87}$Sr atoms into the science trap due to the naturally higher
abundance of  $^{88}$Sr. We use a hold time of 1\,s and we do not use the
532-nm trapping beam. The peak densities for Rb and $^{88}$Sr are $6(3)\times 10^{11}\,$cm$^{-3}$
and $1.1(6)\times 10^{13}\,$cm$^{-3}$.
The temperature of the Rb-Sr mixture before the hold time is 2.2(1)\,$\mu$K.

\textbf{Magnetic field.} We use three pairs of coils to produce a homogeneous
magnetic field across the atomic sample. The primary coils create a field up to
$\sim 500$\,G with a resolution of 40\,mG. These coils are used alone to record
most of the loss features shown in Fig.\ \ref{fig:Fig1_LossFeatures}. A second
pair of coils is employed to resolve the substructure of the [1,0]a and [1,+1]a
loss features, with the primary coils producing bias fields of 278\,G and
294\,G, respectively. The secondary coils create a low magnetic field with a
resolution of 3\,mG. A third pair of coils is used to supplement the primary
coils to observe the [1,-1]c loss feature, creating a bias field of 57\,G.

We calibrate the magnetic field produced by the primary coils up to 290\,G by
spectroscopy on the narrow $^1$S$_0$-$^3$P$_1$ line of $^{88}$Sr. We use a
current transducer (LEM IT 600-S) to interpolate between the calibration points
and to extrapolate to higher fields. We calibrate the secondary coils by
recording one of the three [1,+1]a loss features for different values of the
field from the primary coils.

The magnetic field precision is limited by the resolution and noise of the
power supplies. The inductances of the coils reduce the noise contribution to
less than 40\,mG from all three pairs of coils combined. The accuracy is
limited mainly by drifts in the ambient magnetic field. Monitoring the position
of the $^{88}$Sr MOT and the position of the loss feature due to a known Rb
Feshbach resonance gives an upper bound of 350\,mG for these drifts over the
course of the present work. The calibration and statistical errors are
typically one order of magnitude lower than the drifts. In addition, the
positions of the loss maxima at finite temperature may differ from the
positions of zero-energy Feshbach resonance positions. We account for this
systematic error in the binding energies of Tab.\ \ref{tab:BoundStates} by
adding an uncertainty of 4 times the root-mean-square width of the fitted
Gaussian function.

\section{Author contributions}
V.B., A.C. and L.R. performed the experiments. B.P. and F.S. supervised the
experimental work. P.S.\.{Z}. and J.M.H. contributed theoretical analysis. All
authors were involved in analysis and discussions of the results and
contributed to the preparation of the manuscript.

\end{document}